# LEARN++: Recurrent Dual-Domain Reconstruction Network for Compressed Sensing CT

Yi Zhang, *Senior Member, IEEE*, Hu Chen, Wenjun Xia, Yang Chen, Baodong Liu, Yan Liu, Huaiqiang Sun, and Jiliu Zhou, *Senior Member, IEEE*

*Abstract*—Compressed sensing (CS) computed tomography has been proven to be important for several clinical applications, such as sparse-view computed tomography (CT), digital tomosynthesis and interior tomography. Traditional compressed sensing focuses on the design of handcrafted prior regularizers, which are usually image-dependent and time-consuming. Inspired by recently proposed deep learning-based CT reconstruction models, we extend the state-of-the-art LEARN model to a dual-domain version, dubbed LEARN++. Different from existing iteration unrolling methods, which only involve projection data in the data consistency layer, the proposed LEARN++ model integrates two parallel and interactive subnetworks to perform image restoration and sinogram inpainting operations on both the image and projection domains simultaneously, which can fully explore the latent relations between projection data and reconstructed images. The experimental results demonstrate that the proposed LEARN++ model achieves competitive qualitative and quantitative results compared to several state-of-the-art methods in terms of both artifact reduction and detail preservation.

## I. INTRODUCTION

IN classical computed tomography (CT) image reconstruction theory, a large number of measurements of a target object are required to generate an accurate reconstruction with current commercial analytic algorithms. For clinical applications, the amount of collected data may increase the potential risk of radiation-induced biological diseases and limit the extension of a system to different diagnostic and interventional purposes, such as digital breast tomosynthesis, cardiac CT and cone beam CT (CBCT), for radiation treatment planning [1-3]. Thus, it is of clinical importance to develop an efficient and fast reconstruction method to address the problem of sparse-view CT imaging.

In the past decade, many efforts have been dedicated to the field of compressive sensing (CS) [4, 5], which can effectively address the sparse-view reconstruction problem. The key part of CS is to find a proper *a priori* or sparsifying transform to define the regularization term. Some representative works include the total variation, nonlocal means, wavelets, and low rank. Although these models achieved fruitful results, several critical limitations prevent CS-based methods from achieving clinically practical applications: (i) It is quite difficult to design a universal handcrafted prior regularization for general scanning protocols, which means that there is currently no regularization term that can be applied for all diagnostic purposes. (ii) CS-based reconstruction algorithms are iterative and include multiple projection and backprojection operations. This procedure is time-consuming and especially aggravated by the numerical computation of the complex regularization terms. (iii) The parameter setting is usually image-dependent, and it is almost impossible for radiologists to conduct this tedious tuning process.

Very recently, deep learning (DL) techniques have been introduced into CT imaging and have obtained encouraging results in this field [6-8]. These DL-based methods can partly address the issues mentioned above. In particular, Jin et al. first introduced U-Net for sparse-view CT image restoration [9]. The authors proposed using U-Net followed by a direct inversion operation (for CT, the filtered backprojection (FBP) is applied) to alleviate the artifacts caused by undersampling. Following a similar procedure, in [10], Zhang et al. proposed replacing the convolution layers in U-Net with dense blocks to accelerate training and enhance the expression ability. Aided by the wavelet transform, Kang et al. used a similar network backbone to suppress the noise in low-dose CT images [11]. The same group also sought to bridge the theoretical gaps of DL and classical methods in signal processing and extend the proposed deep convolutional framelets to low-dose and sparse-view CT imaging [12, 13]. Chen et al. introduced the well-known SRCNN model, which was originally designed for image super-resolution, for low-dose CT and further optimized the network architecture using autoencoders and residual connections [14-17]. In [18], the generative adversarial network (GAN) framework was utilized for noise reduction in low-dose CT, and Yang et al. added the Wasserstein distance and perceptual loss to improve the convergence and performance of the GAN [19]. Although methods that directly process images reconstructed by FBP have achieved promising results, as stated by [20], the potential risk of ignoring the consistency to the measurements may lead to unstable and inaccurate results, which are harmful for clinical diagnosis.

On the other hand, there are some studies on DL-based methods that directly process the sinogram. The sparse-view or limited-view problems can be regarded as image inpainting or interpolation in which different networks are designed to recover the missed part in the sinogram [21-24]. However, since the reconstruction is sensitive to the inherent consistency of the sinogram, any improper operations directly on the sinogram may introduce severely unexpected artifacts over an entire reconstructed image. Therefore, incorporating the DL technique into the iterative reconstruction framework is a possible solution for these obstacles. In the classical iterative reconstruction procedure, the data fidelity and image regularization terms are performed alternatively. A direct thought is to substitute the traditional handcrafted regularization term with a trained neural network. Typically, Wu et al. and Wang et al. respectively used the k-sparse autoencoder and RLNet as the constraints, respectively [25, 26]. Since these methods only added network-based regularization terms into the iterative framework, they still suffered from heavy computational costs and laborious parameter adjustment. Another solution does not directly replace the regularization term with a trained network. Inspired by the original idea of unrolling sparse coding [27], many efforts have been made to unroll different optimization algorithms into trainable end-to-end networks. For example, Chen et al. introduced the fields of experts [28] as the regularization term, and the simple gradient decent method was applied to form the iterative numerical scheme [20]. Then, while the number of iterations was fixed, the procedure was unrolled into a network, which enabled the parameters and regularization terms to be trainable with an external dataset. Using a similar idea, in [29], Gupta et al. proposed substituting the projector in a projected gradient descent with a convolutional neural network (CNN). Alder and Öktem replaced the nonlinear proximal operators in the proximal primal-dual optimization method with CNNs [30]. Inspired by the work of ADMM-Net [31, 32], He et al. proposed constraining the reconstruction problem in both data and image domains and unrolled the alternating direction method of multipliers (ADMM)-based optimization method into a network [33]. This kind of method can efficiently accelerate reconstruction and avoid parameter adjustment, but the measurements are used only as data consistency constraints, and the prior information in the data domain is not well explored. Several groups separately proposed adopting two independent networks to process data and images progressively, and these networks can be treated as pseudo dual domain methods [34-38]. However, these methods all filter the sinogram first and then process the reconstructed image later, and this order implicitly involves a certain priority in the reconstruction, which loses the potential interaction between different domains. Furthermore, a pseudoinverse (FBP and the IFFT are used for CT and MRI, respectively) is usually used to connect two domains, which makes these methods essentially single domain oriented.

In this study, we extend our previously proposed network LEARN model to a dual-domain version, dubbed LEARN++. Except for the learned regularization term for the reconstructed image, the data fidelity of LEARN++ model is enriched using an inpainting network, which is also constrained by the measured data. The proposed LEARN++ model can be seen as a cascaded network with successive blocks, which consist of two parallel interactive subnetworks simultaneously handing the sinogram and image in each block. The remainder of this paper is organized as follows. Section II elaborates the details of LEARN++. Section III presents the experimental results and discussions. Section IV concludes this paper.

## II. METHODOLOGY

### A. LEARN

To make this paper self-contained, we first briefly review the main idea of LEARN [20]. Let $\mathbf{x}=(x_1, x_2,...,x_J)^T \in \mathbb{R}^J$ represent a vector of attenuation coefficients. Generally, the CT reconstruction problem estimates $\mathbf{x}$ from $\mathbf{y} = (y_1, y_2,..., y_I)^T \in \mathbb{R}^I$ using measured data after calibration and the log-transform:

$$\mathbf{y} = A\mathbf{x}, \quad (1)$$

where $A \in \mathbb{R}^{I \times J}$ is the projection matrix associated with the scanning geometry. Generally, FBP is applied while the linear system in (1) is well conditioned. Once the sampling procedure is sparse, (1) becomes underdetermined, leading to FBP reconstruction contaminated by severe noise and artifacts, which will heavily affect the clinical diagnosis. Iterative reconstruction that leverages prior knowledge at the expense of significantly increased computational costs is an effective solution for this issue.

Let us consider the iterative reconstruction with a general regularized formulation as

$$\mathbf{x}^* = \arg\min_{\mathbf{x}} \lambda D(\mathbf{x},\mathbf{y}) + R(\mathbf{x}), \quad s.t.\ x_j \geq 0 \quad \forall j, \quad (2)$$

where $D$ represents the data consistency, $R$ represents the regularization term and $\lambda$ denotes the balancing factor. In the field of CS-based methods, representative models include the total variation (TV) and its variants, the tight framelet, dictionary learning, nonlocal means and low-rank matrix factorization [39-47]. In [20], a learning-based model derived using iteration unrolling is given. Specifically, A generalized regularization term, dubbed as fields of experts [28], is introduced into (2) and combined with $\ell_2$ norm-based data

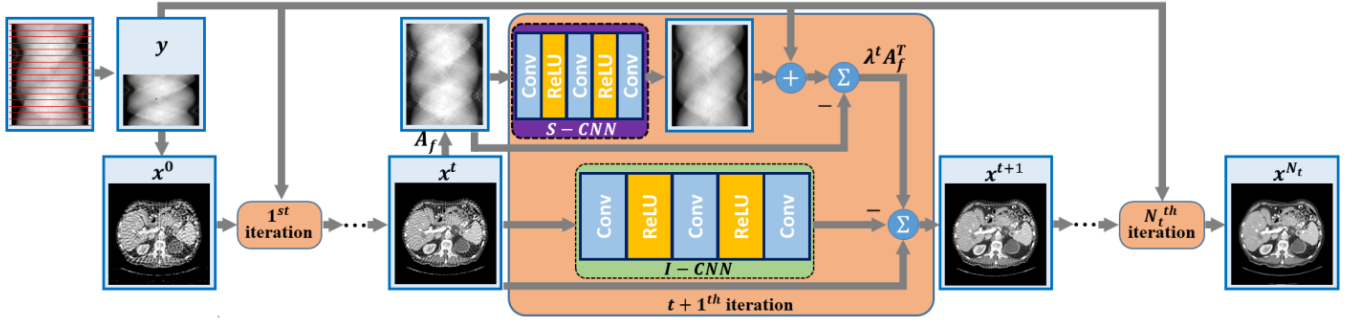

Fig. 1. Overall structure of our proposed LEARN++ network.

fidelity, which can be formulated as

$$\mathbf{x}^* = \arg\min_{\mathbf{x}} \frac{\lambda}{2}\|A\mathbf{x} - \mathbf{y}\|_2^2 + \sum_{k=1}^{K}\phi_k(G_k\mathbf{x}), \quad s.t. \; x_j \geq 0 \; \forall j. \quad (3)$$

where $G_k$ is a transform matrix, $\phi_k(\cdot)$ represents a potential function and $K$ is the number of regularization terms. Assuming that the regularization terms are differentiable, the gradient descent method is applied to (3):

$$\mathbf{x}^{t+1} = \mathbf{x}^t - \left(\lambda A^T(A\mathbf{x}^t - \mathbf{y}) + \sum_{k=1}^{K}(G_k)^T \gamma_k(G_k \mathbf{x}^t)\right), \quad (4)$$

where $T$ denotes the transpose operator and $\gamma_k(\cdot)$ is the derivative of $\phi_k(\cdot)$.

(4) can be converted to an iteration-dependent version, and the regularization terms in (4) can be substituted by a convolution module to learn the latent representation of training samples as

$$\mathbf{x}^{t+1} = \mathbf{x}^t - \left(\lambda^t A^T(A\mathbf{x}^t - \mathbf{y}) + f_{I-CNN}(\mathbf{x}^t, \Theta_1^t)\right), \quad (5)$$

where $\lambda^t$ is the iteration-dependent parameter, which may differ in each iteration. $f_{I-CNN}(\cdot)$ represents the convolution module, and $\Theta_1$ refers to the learned network parameters. Thus, the iteration procedure of (4) can be unrolled into a neural network with a fixed number of iterations. The iteration-dependent regularization terms and balancing parameter can be learned using existing samples. The computation of $\lambda A^T(A\mathbf{x}^t - \mathbf{y})$ can be treated as a data consistency layer, and the whole block has a skip connection, which is similar to the residual block in [48].

### B. LEARN++

In (5), CNN-based filtering is only applied to the intermediate results in the image domain, and there is no additional operation in the projection domain. Recently, some studies have demonstrated that processing data in both data and image domains can effectively improve the reconstruction performance. However, existing dual-domain methods for CT reconstruction handle projection and image data sequentially, which potentially impose an implicit prior on the reconstruction procedure and probably neglects the mutual interaction between two domains [34, 35, 37]. In this paper, to incorporate the power of dual-domain methods, based on the LEARN framework, we propose a parallel interactive dual-domain network for CT reconstruction, dubbed LEARN++.

In the proposed LEARN++, we focus on the data consistency layer $\lambda A^T(A\mathbf{x}^t - \mathbf{y})$ in (5). In this term, the intermediate result $\mathbf{x}^t$ is projected to the projection domain to calculate the residual with measured data. Then, the residual is backprojected to the image domain and weighted together with the CNN filtered result to form the final output of one iteration block. For the sparse sampling problem, $A$ and $A^T$ correspond to the sparse sampling projection and backprojection operators, respectively, which only consider the impact of measured data. Since the sampling rate is low, limited information from the intermediate image is utilized to estimate the residual, which may cause more data errors and have a negative impact on the convergence speed [33, 35, 36]. To overcome this deficiency of the original LEARN, we introduce prior knowledge of the full-sampling projection matrix $A_f$ and CNN-based sinogram inpainting. Let $P=\lambda A^T(A\mathbf{x}^t - \mathbf{y})$ in (5). Then, we replace $P$ with $P'$, which is defined as

$$P' = \lambda^t A_f^T(A_f \mathbf{x}^t - f_{S-CNN}(A_f \mathbf{x}^t, \Theta_2^t) \oplus \mathbf{y}), \quad (6)$$

where $f_{S-CNN}(\cdot)$ represents the convolution module for projection data, which plays a similar role in DL-based sinogram inpainting methods [21-24]; and $\Theta_2$ denotes the learned parameter set for $f_{S-CNN}(\cdot)$. $f_{S-CNN}(\cdot)$ outputs the processed full-sampling projection data $\mathbf{y}_f^t$ and $\oplus$ represents overwriting the corresponding indexes in $\mathbf{y}_f^t$ with $\mathbf{y}$ and the other indexes remain the same. Comparing $P$ with $P'$, it can be seen that two critical modifications are applied: 1) the undersampling projection matrix $A$ is substituted by $A_f$, which is dedicated to imposing the constraint of both measured and estimated data from the previous iteration block; and 2) a CNN module is employed to filter the estimated data, and then the original part $\mathbf{y}$ is inserted into the filtered projection data. Then, (5) is converted into

$$\mathbf{x}^{t+1} = \mathbf{x}^t - \lambda^t A_f^T(A_f \mathbf{x}^t - f_{S-CNN}(A_f \mathbf{x}^t, \Theta_2^t) \oplus \mathbf{y}) \\ - f_{I-CNN}(\mathbf{x}^t, \Theta_1^t). \quad (7)$$

By unrolling (7) with a fixed number of iterations $N_t$, we obtain the network architecture of LEARN++, which is depicted in Fig. 1. In Fig. 1, one iteration block is composed of three parallel submodules, including a residual connection, an image domain subnetwork and a projection domain subnetwork. For simplicity, similar to LEARN, a three-layer CNN is utilized as the image domain subnetwork. Meanwhile, the same structure is adopted for the projection domain subnetwork. It is worth mentioning that the utilization of more advanced techniques, such as dense connection or attention mechanism, may further improve the model's performance, but this is beyond the scope of this paper, which is dedicated to demonstrating the advantage of integrating the powers of dual domain filtering.

The proposed LEARN++ network is trained in a supervised manner, which requires paired training samples, including the sparse sampling projection data and their corresponding high-quality images. The training set $D$ contains $S$ paired samples $(\mathbf{x}_s, \mathbf{y}_s)_{s=1}^S$.

The loss function of LEARN++ consists of three components as follows.

(1) The mean squared error (MSE) $L_{I-MSE}$ is used to measure the reconstruction error in the image domain as

$$L_{I-MSE} = \frac{1}{S}\sum_{s=1}^{S} \|\hat{\mathbf{x}}_s - \mathbf{x}_s\|_2^2, \qquad (8)$$

where $\hat{\mathbf{x}}$ denotes the predicted reconstruction result.

(2) The MSE $L_{S-MSE}$ is utilized to measure the reconstruction error in the projection domain as

$$L_{S-MSE} = \frac{1}{S}\sum_{s=1}^{S} \|A_f \hat{\mathbf{x}}_s - A_f \mathbf{x}_s\|_2^2. \qquad (9)$$

(3) The perceptual loss, which measures the differences in a specific feature space, has been proven to be effective at recovering details more consistent with the radiologist's reading habits [19, 49]. Following the idea of [19], a pretrained VGG-19 was chosen, and the feature map extracted from the 16th convolutional layer was adopted to define the loss as

$$L_{VGG} = \frac{1}{S}\sum_{s=1}^{S} \|\varphi(\hat{\mathbf{x}}) - \varphi(\mathbf{x})\|_2^2, \qquad (10)$$

where $\phi(\cdot)$ represents the feature extraction function.

Then, the total loss function of the proposed LEARN++ model is formulated as

$$L = L_{I-MSE} + \lambda_1 L_{S-MSE} + \lambda_2 L_{VGG}. \qquad (11)$$

In this paper, the loss function was optimized using Adam [50].

## III. RESULTS AND DISCUSSION

The public dataset "*the 2016 NIH-AAPM-Mayo Clinic Low Dose CT Grand Challenge*", which was established and authorized by the Mayo Clinic, was used in our experiments to evaluate the performance of the proposed LEARN++. This dataset contains 5,936 routine dose human torso slices with a 1 mm thickness from 10 patients. To reconstruct the reference image, the projection data were simulated in fan-beam geometry from 512 views per scan with Siddon's ray-driven algorithm [51]. We simply downsampled the projection data to 128 and 64 views to simulate the sparse-view cases with sampling rates of 1/4 and 1/8, respectively. To build the dataset for our study, 25 slices for each patient were randomly selected, and the dataset was composed of a total of 250 images. Some representative slices are given in Fig. 2. In the training stage, 200 images from 8 patients were chosen as the training set, and the remaining 50 images from the remaining 2 patients were used as the testing set. For the first random selection, 37 thoracic, 82 abdominal and 81 pelvic slices were entered into the training set, and the testing dataset consisted of 8 thoracic, 18 abdominal and 24 pelvic slices. In addition, cross validation was performed to evaluate the generalization of the proposed model.

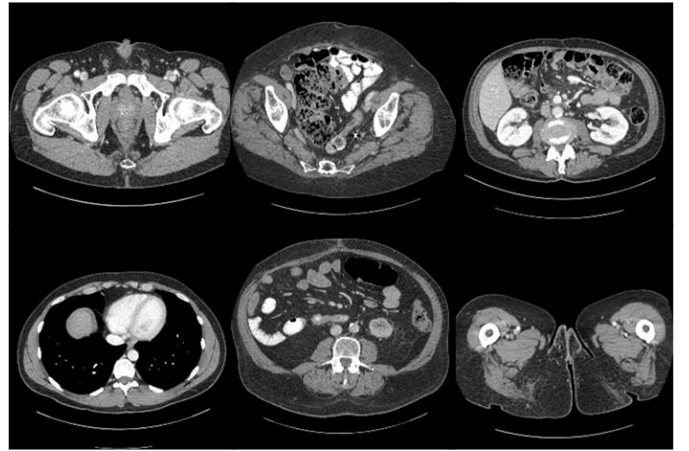

Fig. 2. Examples in the dataset. The display window is [-150 250] HU.

In the experiments, the performances of several state-of-the-art DL-based approaches, including FBPConvNet [9], WGAN-VGG [19], DD-Net [10], LEARN [20] and DP-ResNet [35], were evaluated and compared. The first three methods were recently proposed postprocessing network models. FBPConvNet adopts U-Net as the backbone of the network. WGAN-VGG introduces the WGAN as its basic framework; and the perceptual loss, which is implemented by the VGG network, is added into the loss function to preserve more details. DD-Net combines DenseNet and deconvolution and applies residual connections to enhance the expression ability of the network and accelerate the training procedure. LEARN is an unrolling iteration network for sparse-view CT reconstruction. DP-ResNet is a dual-domain network for CT image reconstruction. It successively processes the input measured data in the projection and image domains, and FBP is used to join two subnetworks. The source codes of all these network models were authorized and provided by the authors, and we retrained these models with our dataset. All the network and training parameters of the compared methods were the same. Specifically, since postprocessing methods (FBPConvNet, WGAN-VGG and DD-Net) require more training data to guarantee the model performance, a total of 4800 slices, including the 200 images in LEARN++'s training set, were chosen. The peak signal-to-noise ratio (PSNR) and structural similarity index measure (SSIM) were utilized as the quantitative metrics of image quality.

In the experiments, the network parameters of LEARN++ were set consistent with those of LEARN to provide a fair comparison. The number of iteration blocks $N_t$ was set to 50, and the number of filters in each iteration block was set to 48 for the first two layers and 1 for the last layer. The kernel size of these filters was set to $5 \times 5$. $\lambda^t$ was initialized to 0, and the weights in the kernels were initialized using the Xavier method [52]. The learning rate was initially set to $10^{-4}$ and decreased by 10 times every 100 epochs to $10^{-6}$. The proposed LEARN++ model was implemented with TensorFlow on a GPU station (Intel Xeon E5-2603 v4 CPU, 128 GB RAM and Tesla P100 GPU*2). The codes for this work are available at https://github.com/maybe198376/LEARN-Plus-Plus.

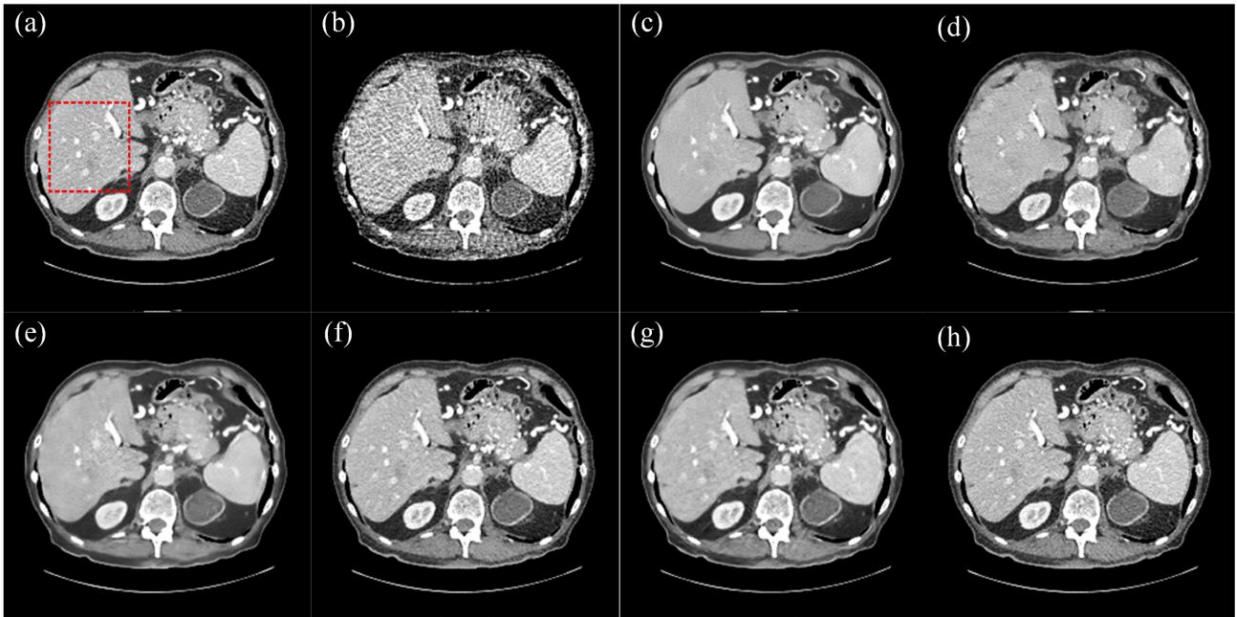

Fig. 3. Abdomen images reconstructed using different methods with 128 views. (a) The reference image, (b) FBP, (c) FBPConvNet, (d) WGAN-VGG, (e) DD-Net, (f) LEARN, (g) DP-ResNet, and (h) LEARN++. The red dotted box marks an ROI, which is enlarged in Fig. 4. The display window is [−150 250] HU.

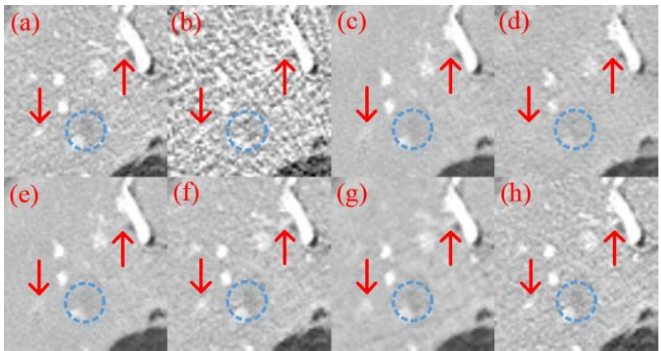

Fig. 4. Magnified ROI labeled by the red dotted box in Fig. 3(a). (a) The reference image, (b) FBP, (c) FBPConvNet, (d) WGAN-VGG, (e) DD-Net, (f) LEARN, (g) DP-ResNet, and (h) LEARN++. The red arrows indicate some contrast-enhanced vessels, which can only be well recognized by LEARN++. The blue dotted circle indicates the low-contrast metastasis. The display window is [−150 250] HU.

### A. Visual Results

To demonstrate the performance of the proposed LEARN++ model, two typical slices are chosen. Fig. 3 demonstrates an abdomen slice reconstructed using different methods with 128 views. It can be observed in Fig. 3(b) that the result reconstructed by FBP is severely degraded by streak-like artifacts, which suppress the most clinically important details. The metastasis in the red dotted box is hard to identify via visual inspection. In Figs. 3(c)-(h), although most artifacts are eliminated, the details in the results of FBPConvNet, WGAN-VGG, and DD-Net are blurred to varying degrees. Specifically, some contrast-enhanced vessels in the lung, heart and spleen regions are distorted or even oversmoothed. To further visualize the improvement of our method over other methods, Fig. 4 shows the magnified region of interest (ROI) labeled by the red dotted box. Two red arrows indicate two contrast-enhanced vessels, as shown in Fig. 4(a). In the FBPConvNet, WGAN-VGG, and DD-Net results, these structures almost disappear. In Fig. 4(g), DP-ResNet slightly blurs these vessels, and artifacts are still visible in the middle region. These details can be well recognized in Figs. 4(f) and (h), and LEARN++ obtains the closest visual effect to the reference image. The low-contrast metastasis marked by a blue circle is well reconstructed by LEARN++, and it has a similar contrast to the one in the reference image. Fig. 5 gives the absolute difference images related to the reference image to further illustrate the merits of LEARN++. It is clearly noticed that LEARN++ yields the smallest residual to the reference image, which can be treated as another powerful evidence of its artifact suppression and detail preservation abilities.

Fig. 6 shows another reconstructed abdomen image using different methods with 64 views. As the number of sampling views further continues to decrease, the FBP results become even worse. Due to the severe noise and artifacts, it is impossible to obtain any useful information in Fig. 6(a). All DL-based methods suppress the artifacts to varying degrees. Two red arrows point to the liver and spleen regions, which can distinguish the visual effects of different methods. DD-Net and DP-ResNet cannot remove the streak artifacts in the spleen well, and the details in the liver are also eroded. WGAN-VGG achieves better artifact reduction performance, but the details, including the vessels and mottle-like structures, are not

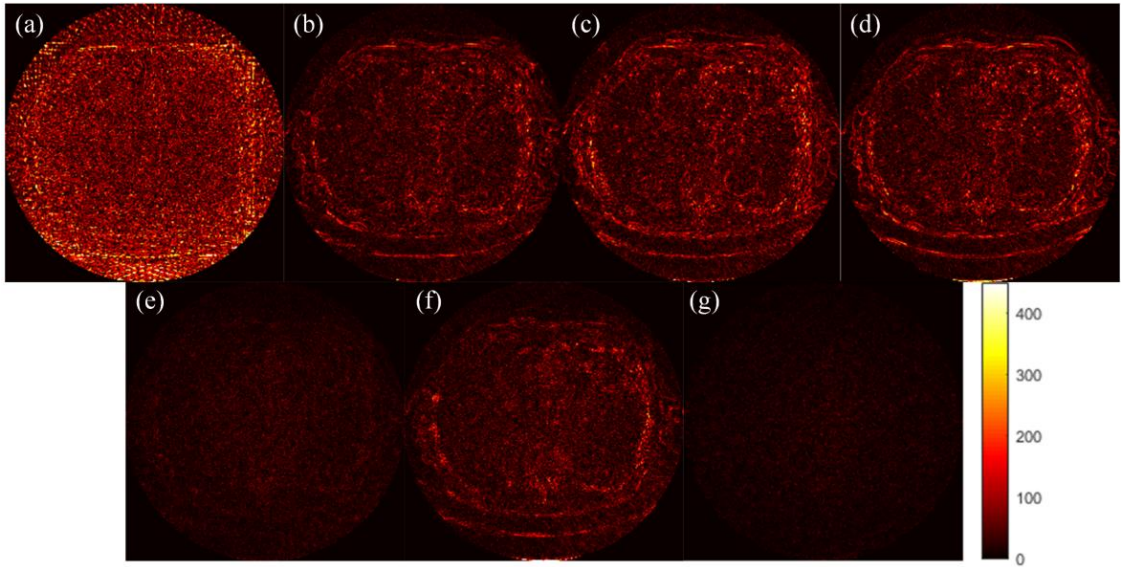

Fig. 5. Absolute difference images related to the reference image. (a) FBP, (b) FBPConvNet, (c) WGAN-VGG, (d) DD-Net, (e) LEARN, (f) DP-ResNet, and (g) LEARN++.

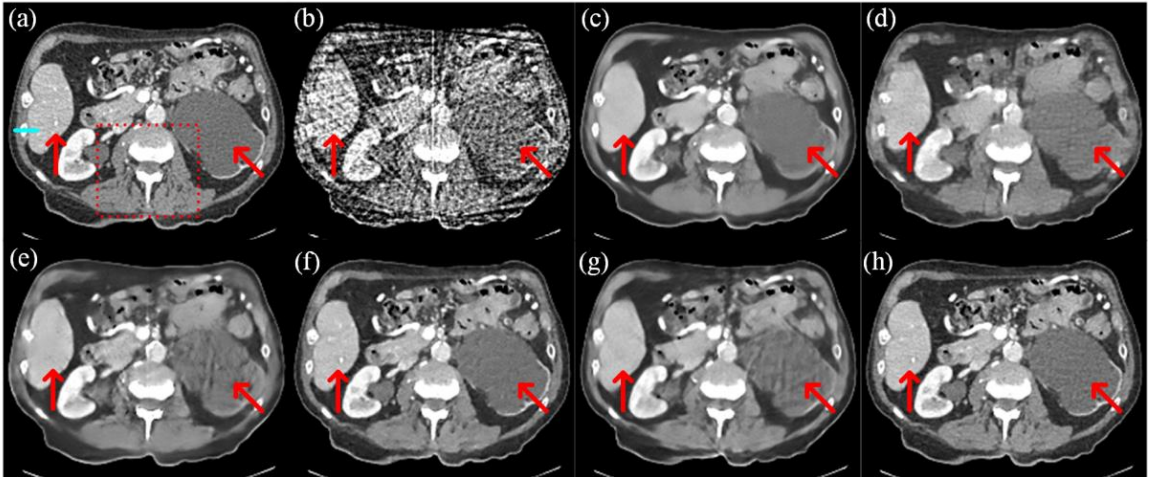

Fig. 6. Abdomen images reconstructed using different methods with 64 views. (a) The reference image, (b) FBP, (c) FBPConvNet, (d) WGAN-VGG, (e) DD-Net, (f) LEARN, (g) DP-ResNet, and (h) LEARN++. The arrows indicate two regions containing details recovered differently by the different methods. The red dotted box marks an ROI, which is magnified in Fig. 7. The profiles along the blue line are plotted in Fig. 8. The display window is [−150 250] HU.

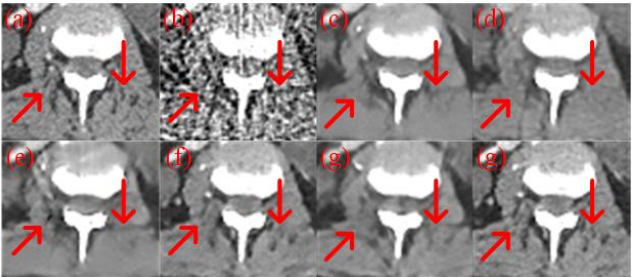

Fig. 7. Magnified ROI labeled by red dotted box in Fig. 6(a). (a) The reference image, (b) FBP, (c) FBPConvNet, (d) WGAN-VGG, (e) DD-Net, (f) LEARN, (g) DP-ResNet, and (h) LEARN++. The arrows indicate two regions containing features recovered differently by the competing methods. The display window is [−150 250] HU.

Table I Quantitative results of the reconstructed images using both 128 and 64 views in Fig. 3

| No. of views | 128 | | 64 | |
| --- | --- | --- | --- | --- |
| | PSNR | SSIM | PSNR | SSIM |
| FBPConvNet | 37.02 | 0.9313 | 33.74 | 0.9078 |
| WGAN-VGG | 35.36 | 0.9070 | 32.09 | 0.8748 |
| DD-Net | 35.75 | 0.9186 | 32.46 | 0.8931 |
| LEARN | 42.76 | 0.9744 | 39.66 | 0.9659 |
| DP-ResNet | 35.75 | 0.9398 | 32.40 | 0.9107 |
| LEARN++ | **46.71** | **0.9885** | **40.75** | **0.9711** |

recovered. LEARN and LEARN++ seem to achieve the best visual effects. To further validate the improvement of the proposed LEARN++ compared to other methods, Fig. 7 illustrates the magnified regions indicated by the red dotted box in Fig. 6(a). Some structural details near the spine are marked by red arrows. The figure shows that LEARN++ achieves the best performance and has the clearest details compared to the other methods. To compare the detail preservation abilities of different methods, a horizontal profile, which is marked with a blue line in Fig. 6(a), is plotted in Fig. 8. Two green arrows point to two locations with obvious visual differences. It is easy to observe that LEARN++ obtains the most consistent result with respect to the reference image compared to all the other

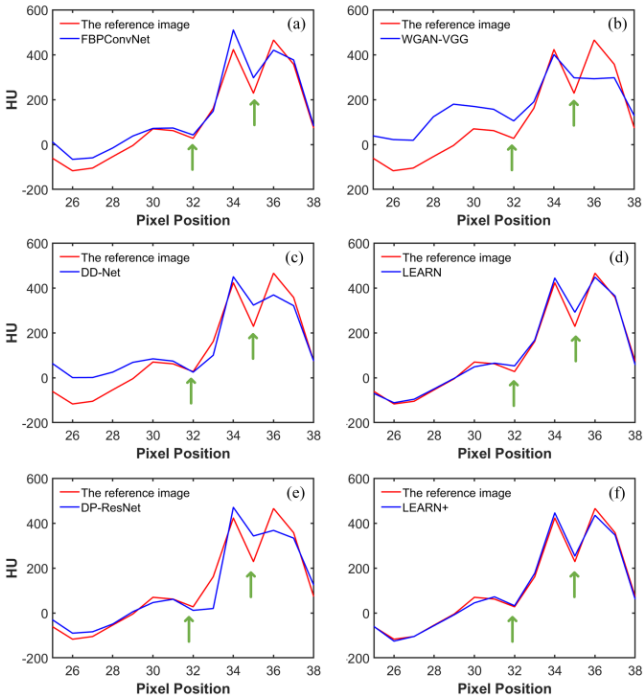

Fig. 8. Horizontal profiles along the blue line located at the 139th row and 24th to 38th columns in Fig. 6(a) of the reference image versus the images reconstructed using different methods. (a) FBPConvNet, (b) WGAN-VGG, (c) DD-Net, (d) LEARN, (e) DP-ResNet, and (f) LEARN++. Two green arrows indicate two locations with obvious visual differences.

methods.

*B. Quantitative and Qualitative Results*

Table I lists the quantitative results of the reconstructed images using both 128 and 64 views in Fig. 3. The table shows that our proposed LEARN++ model achieves the best scores in both metrics and makes noticeable improvements over LEARN, which is consistent with the visual inspection in Fig. 3.

The quantitative results of the reconstructed images using both 128 and 64 views in Fig. 6 are given in Table II. The results have a similar trend as those in Table I in that LEARN++ obtains better PSNR and SSIM scores than all the other methods.

Table III summarizes the quantitative results for the full cross validation by calculating the means of both metrics on the test set using 128 and 64 views for all methods. The results confirm the advantages of the proposed LEARN++ model and the effectiveness of the dual-domain network. LEARN and LEARN++, which include the projection correction operation in the network, have better performance than the compared postprocessing methods. Furthermore, the introduction of the dual-domain network can improve the reconstruction, which is supported by the observation that DP-ResNet obtains better results than FBPConvNet, WGAN-VGG and DD-Net and LEARN++ obtain better results than LEARN.

Table II Quantitative results of the reconstructed images using both 128 and 64 views in Fig. 6

| No. of views | 128 | | 64 | |
|---|---|---|---|---|
| | PSNR | SSIM | PSNR | SSIM |
| FBPConvNet | 38.28 | 0.9313 | 34.26 | 0.8977 |
| WGAN-VGG | 36.50 | 0.9149 | 32.69 | 0.8751 |
| DD-Net | 36.70 | 0.9186 | 32.70 | 0.8806 |
| LEARN | 44.72 | 0.9819 | 39.41 | 0.9522 |
| DP-ResNet | 36.93 | 0.9554 | 33.01 | 0.9045 |
| LEARN++ | **48.86** | **0.9918** | **40.40** | **0.9600** |

Table III Quantitative results of the reconstructed images using both 128 and 64 views on the full cross validation

| No. of views | 128 | | 64 | |
|---|---|---|---|---|
| | PSNR | SSIM | PSNR | SSIM |
| FBPConvNet | 37.70 | 0.9431 | 34.56 | 0.9051 |
| WGAN-VGG | 35.97 | 0.9178 | 33.34 | 0.8788 |
| DD-Net | 36.12 | 0.9297 | 33.04 | 0.8815 |
| LEARN | 44.78 | 0.9835 | 40.73 | 0.9660 |
| DP-ResNet | 37.10 | 0.9580 | 33.71 | 0.9137 |
| LEARN++ | **49.03** | **0.9924** | **41.93** | **0.9724** |

For the qualitative evaluation, 20 images were randomly selected from the results of the cross validation with 64 views using different methods for an image reader study. Two radiologists with ten and eight years of clinical experience, respectively, assessed these images independently. Noise suppression, artifact reduction, contrast retention and overall quality were employed as four subjective indicators on a five-point scale, where 1=unacceptable and 5=excellent. The reference images were treated as the gold standard. For the test set, the scores are provided as means±SDs (average scores±standard deviations of the two radiologists), and Student's t-test was performed to statistically evaluate the significance of the differences. The results are shown in Table IV. The table shows that both radiologists suggest the FBP results are unacceptable in terms of each indicator. All the methods can improve the results with a significant promotion in each score. LEARN and LEARN++ occupy the top two positions, and LEARN++ obtains slightly better results than LEARN. Student's t-test demonstrates a coherent trend that LEARN and LEARN++ do not have statistically significant differences in all qualitative indicators to the reference image.

Table IV Statistical analysis of subjective quality scores using different algorithms with 64 views (Mean ± SDs)

| | Reference | FBP | FBPConvNet | WGAN-VGG | DD-Net | LEARN | DP-ResNet | LEARN++ |
|---|---|---|---|---|---|---|---|---|
| Noise suppression | 3.88±0.31 | 1.13±0.12* | 3.43±0.32 | 3.08±0.69* | 3.12±0.48* | 3.52±0.42 | 3.38±0.46 | **3.60±0.38** |
| Artifact reduction | 4.1±0.33 | 1.08±0.08* | 3.37±0.34* | 2.85±0.50* | 3.01±0.33* | 3.56±0.27 | 3.12±0.37* | **3.58±0.35** |
| Contrast retention | 4.23±0.22 | 1.66±0.24* | 3.48±0.46* | 3.41±0.34* | 3.43±0.56* | 3.52±0.33 | 3.50±0.23 | **3.53±0.28** |
| Detail preservation | 3.83±0.41 | 1.08±0.08* | 2.37±0.78* | 2.42±0.88* | 2.39±0.76* | 3.22±0.46 | 2.71±0.67* | **3.49±0.53** |
| Overall quality | 3.85±0.30 | 1.22±0.17* | 2.54±0.56* | 2.46±0.41* | 2.42±0.79* | 3.29±0.35 | 2.65±0.56* | **3.42±0.31** |

* indicates $p < 0.05$, which means significantly different.

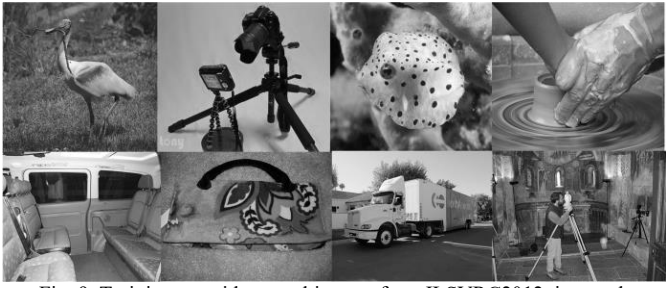

Fig. 9. Training set with natural images from ILSVRC2012_img_val.

*C. Analytical Studies*

In this section, some analytical studies were performed to investigate the impacts of some related factors on the proposed reconstruction network. The same dataset and training parameters used in earlier experiments were adopted.

*1) Main Components of the Loss Function*

In our LEARN++, the proposed loss function contains three components, including the reconstruction loss in the image domain $L_{I-MSE}$, the reconstruction loss in the projection domain $L_{S-MSE}$ and the perceptual loss $L_{VGG}$. The quantitative gain of each loss component is given in Table V. It can be observed that introducing the loss in the projection domain can enhance the performance; and as the perceptual loss is added, the scores are further improved.

*2) Network Trained with Natural Images*

In practice, it is difficult to obtain a sufficient number of well-labeled samples for CT reconstruction. A possible way to alleviate this problem is to train the network with natural images. In this section, we evaluate the performance of our proposed network trained with natural images (LEARN++_n). The same network trained with CT images was compared as the baseline (LEARN++_o). The natural images randomly selected from ILSVRC2012_img_val, which is part of ImageNet, were grayed and normalized to the same range and size as the CT images. Fig. 9 illustrates several examples in the natural image dataset. The projection data of natural images were simulated with the same scanning geometry as the CT images. Four typical slices are chosen, and the visual results and quantitative scores are demonstrated in Fig. 10. In Fig. 10, the images in the first row were reconstructed by LEARN++_o, and the images in the second row were reconstructed by LEARN++_n. The figure shows that the results reconstructed from both datasets are quite close. Especially for the results of 128 views, it is hard to identify any conspicuous differences. For the 64 view case, the results of LEARN++_n look slightly smoother than those of LEARN++_o. The quantitative scores confirm this observation that when the number of sampling views is large, the results reconstructed from both CT and natural images are similar; and when the number is small, the results of LEARN++_n will decay faster than LEARN++_o.

*3) Semisupervised Learning*

Since well-labeled data are difficult to collect in clinical practice, there is great potential for semisupervised or unsupervised learning in the medical imaging field. In this subsection, we proposed a simple implementation of semisupervised learning for our proposed LEARN++. We replaced part of the training samples with some samples without labels, which means that only a portion of the undersampled projection data have corresponding images reconstructed with full-sampled projection data. For the part of training samples without labels, the following loss function is used:

$$L_{S-MSE-U} = \frac{1}{S_U} \sum_{s=1}^{S_U} \|A\hat{\mathbf{x}}_s - \mathbf{y}\|_2^2, \quad (12)$$

where $S_U$ represents the number of unlabeled samples. It can be seen that no label is involved in (12). In the training stage, for the dataset composed of both labeled and unlabeled samples, (11) is adopted when training using labeled samples, and (12) is applied when training using unlabeled samples. To validate the performance of our proposed semisupervised learning strategy, four LEARN++ models, including LEARN++_0, LEARN++_5, LEARN++_10 and LEARN++_100, were trained with 0%, 5%, 10% and 100% labeled samples in the 64 view dataset, respectively. The quantitative results are summarized in Table VI. The table shows that as the proportion of training samples with labels decreases, the scores begin to decline, but the performance decay is not very remarkable. Even for the situation in which 0% labeled data are used, which can be treated as unsupervised learning, the scores are still higher than those of the postprocessing methods reported in Table III. Two typical slices reconstructed by LEARN++ trained using the datasets with different proportions of labeled data are illustrated in Fig. 11. Both cases show that LEARN++_100 obtained only slightly better visual effects than other models, which supports the scores in Table VI.

Table V. Quantitative gains obtained by the different loss components

| Components of loss function | | | 64 views | | 128 views | |
|---|---|---|---|---|---|---|
| $L_{I-MSE}$ | $L_{S-MSE}$ | $L_{VGG}$ | PSNR | SSIM | PSNR | SSIM |
| √ | | | 41.06 | 0.9655 | 48.50 | 0.9824 |
| √ | √ | | 41.53 | 0.9685 | 48.90 | 0.9897 |
| √ | √ | √ | **41.93** | **0.9724** | **49.03** | **0.9924** |

Table VI. Quantitative results of the reconstructed images using different proportions of labeled data

| | PSNR | SSIM |
|---|---|---|
| LEARN++_0 | 38.83 | 0.9503 |
| LEARN++_5 | 39.02 | 0.9519 |
| LEARN++_10 | 39.06 | 0.9527 |
| LEARN++_100 | **41.93** | **0.9724** |

## IV. DISCUSSION AND CONCLUSION

In this paper, LEARN++ is proposed to extend the original LEARN model to a dual-domain version. LEARN unrolls the iterative algorithm into a network and learns the regularization terms as convolutional kernels. However, the measured data are only considered as a data consistency term, and no extra operation is applied to them. Actually, the sparse-view problem can be treated as artifact reduction in the image domain or super-resolution in the projection domain. Current methods usually either consider this problem separately or sequentially, which ignores the information interaction between two domains. The proposed LEARN++ model integrates two parallel and interactive subnetworks to perform image restoration and sinogram inpainting operations on both domains simultaneously, which can fully explore the latent relation between projection data and reconstructed images. In the experiments, the proposed LEARN++ model demonstrates competitive performance in terms of both artifact reduction and detail preservation compared to several state-of-the-art reconstruction methods, including FBPConvNet, WGAN-VGG, DD-Net, LEARN and DP-ResNet.

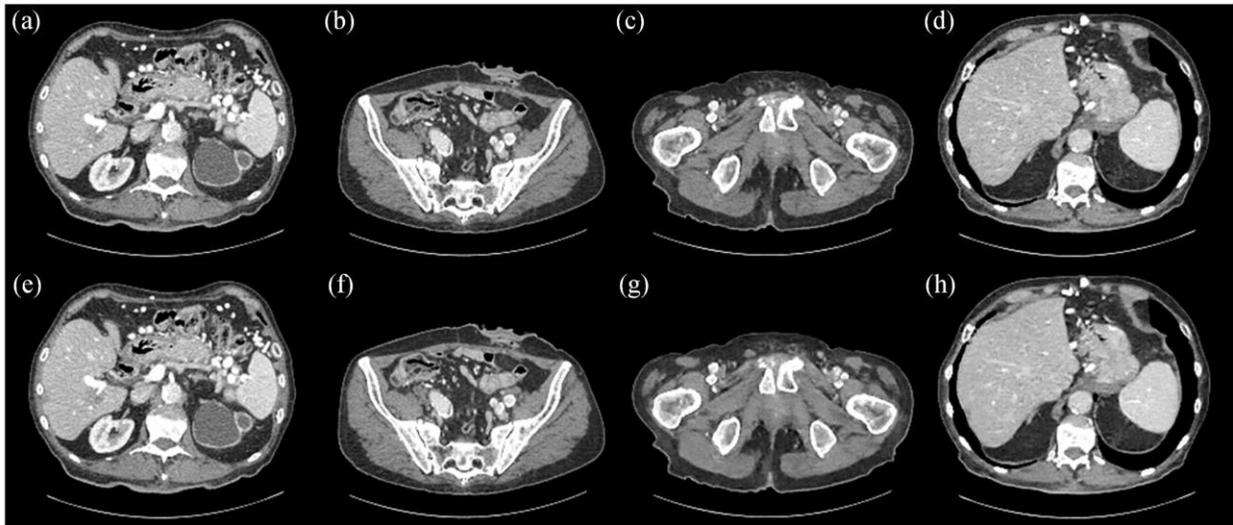

Fig. 10. Results reconstructed by LEARN++ trained with different training sets. The first and second columns were reconstructed with 128 views. The third and fourth columns were reconstructed with 64 views. (a)-(d) were reconstructed using LEARN++_o trained with CT images and (e)-(h) were reconstructed using LEARN++_n. PSNR/SSIM: (a) 47.60/0.9904, (b) 49.44/0.9935, (c) 43.43/0.9777, (d)43.16/0.9784, (e) 47.74/0.9907, (f) 49.56/0.9935, (g) 40.89/0.9629, and (h) 41.84/0.9719.

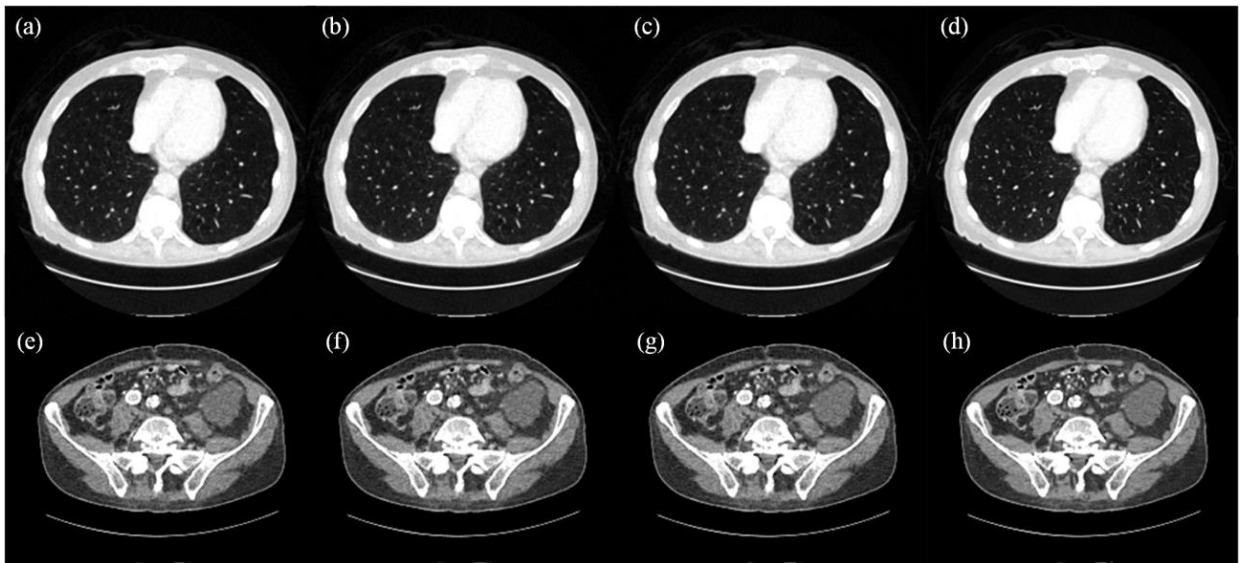

Fig. 11. Two typical slices reconstructed by LEARN++ trained using the datasets with different proportions of labeled data. The first to fourth columns were reconstructed using LEARN++_0, LEARN++_5, LEARN++_10 and LEARN++_100. PSNR/SSIM: (a) 36.76/0.9360, (b) 36.89/0.9377, (c) 36.81/0.9375, (d)40.16/0.9681, (e) 39.24/0.9548, (f) 39.44/0.9565, (g) 39.52/0.9575, and (h) 42.39/0.9769.

It is worth mentioning that similar to other studies based on iteration unrolling, multiple models need to be trained for different scanning protocols, which limits the clinical deployment of our proposed model. A possible solution is to include the corresponding scanning protocol as auxiliary input with projection data [53]. Another benefit accompanied with this solution is that we do not need to store the system matrix in the memory during the training and inference stages.

In future work, more efforts will be made to solve the issue mentioned above, and more datasets and sampling strategies will be included for validation. In addition, some unsupervised learning techniques such as disentangled representation and invertible neural networks will be considered.

## References

[1] S. Schafer et al., "Mobile C-arm cone-beam CT for guidance of spine surgery: Image quality, radiation dose, and integration with interventional guidance," *Med. Phys.*, vol. 38, no. 8, pp. 4563–4574, 2011.

[2] P. FitzGerald et al., "Quest for the ultimate cardiac CT scanner," *Med. Phys.*, vol. 44, no. 9, pp. 4506–4524, 2017.

[3] J. Shan et al., "Stationary chest tomosynthesis using a CNT X-ray source array," *Proc. SPIE*, vol. 8668, p. 86680E, 2013.
[4] D. L. Donoho, "Compressed sensing," *IEEE Trans. Inf. Theory*, vol. 52, no. 4, pp. 1289–1306, Apr. 2006.
[5] E. J. Cand`es, J. Romberg, and T. Tao, "Robust uncertainty principles: Exact signal reconstruction from highly incomplete frequency information," *IEEE Trans. Inf. Theory*, vol. 52, no. 2, pp. 489–509, Feb. 2006.
[6] G. Wang, "A perspective on deep imaging," *IEEE Access*, vol. 4, pp. 8914–8924, 2016.
[7] G. Wang, J. C. Ye, K. Mueller, and J. A. Fessler, "Image reconstruction is a new frontier of machine learning," *IEEE Trans. Med. Imag.*, vol. 37, no. 6, pp. 1289–1296, Jun. 2018.
[8] G. Wang, Y. Zhang, X. Ye, and X. Mou, *Machine Learning for Tomographic Imaging*. Bristol: IOP Publishing, 2020.
[9] K. H. Jin, M. T. McCann, E. Froustey, and M. Unser, "Deep convolutional neural network for inverse problems in imaging," *IEEE Trans. Image Process.*, vol. 26, no. 9, pp. 4509–4522, Sep. 2017.
[10] Z. Zhang, X. Liang, X. Dong, Y. Xie, and G. Cao, "A sparse-view CT reconstruction method based on combination of DenseNet and deconvolution," *IEEE Trans. Med. Imag.*, vol. 37, no. 6, pp. 1407–1417, Jun. 2018.
[11] E. Kang, J. Min, and J. C. Ye, "A deep convolutional neural network using directional wavelets for low-dose X-ray CT reconstruction," *Med. Phys.*, vol. 44, no. 10, pp. e360–e375, 2017.
[12] E. Kang, W. Chang, J. Yoo, and J. C. Ye, "Deep convolutional framelet denoising for low-dose CT via wavelet residual network," *IEEE Trans. Med. Imag.*, vol. 37, no. 6, pp. 1358–1369, Jun. 2018.
[13] Y. Han and J. C. Ye, "Framing U-Net via deep convolutional framelets: application to sparse-view CT," *IEEE Trans. Med. Imag.*, vol. 37, no. 6, pp. 1418–1429, Jun. 2018.
[14] H. Chen et al., "Low-dose CT via convolutional neural network," *Biomed. Opt. Exp.*, vol. 8, no. 2, pp. 679–694, 2017.
[15] H. Chen et al., "Low-dose CT with a residual encoder-decoder convolutional neural network," *IEEE Trans. Med. Imag.*, vol. 36, no. 12, pp. 2524–2535, Dec. 2017.
[16] Y. Liu and Y. Zhang, "Low-dose CT restoration via stacked sparse denoising autoencoders," *Neurocomputing*, vol. 284, no. 1, pp. 80–89, 2018.
[17] W. Du, H. Chen, P. Liao, H. Yang, G. Wang, and Y. Zhang, "Visual attention network for low-dose CT," *IEEE Signal Process. Let.*, vol. 26, no. 8, pp. 1152–1156, 2019.
[18] J. M. Wolterink, T. Leiner, M. A. Viergever, and I. Išgum, "Generative adversarial networks for noise reduction in low-dose CT," *IEEE Trans. Med. Imag.*, vol. 36, no. 12, pp. 2536–2545, Dec. 2017.
[19] Q. Yang et al., "Low-dose CT image Denoising using a generative adversarial network with Wasserstein distance and perceptual loss," *IEEE Trans. Med. Imag.*, vol. 37, no. 6, pp. 1348–1357, Jun. 2018.
[20] H. Chen et al., "LEARN: Learned experts' assessment-based reconstruction network for sparse-data CT," *IEEE Trans. Med. Imag.*, vol. 37, no. 6, pp. 1333–1347, Jun. 2018.
[21] Z. Li, et al., "A sinogram inpainting method based on generative adversarial network for limited-angle computed tomography," in *Proc. Fully Three-Dimensional Image Reconstruction Radiol. Nucl. Med. (Fully3D)*, 2019, 1107220.
[22] J. Gu and J. C. Ye, "Multi-scale wavelet domain residual learning for limited-angle CT reconstruction," in *Proc. Fully Three-Dimensional Image Reconstruction Radiol. Nucl. Med. (Fully3D)*, 2017, pp. 443x447.
[23] H. Lee, J. Lee, H. Kim, B. Cho, and S. Cho, "Deep-neural-network-based sinogram synthesis for sparse-view CT image reconstruction," *IEEE Trans. Radiat. Plasma Med. Sci.*, vol. 3, no. 2, pp. 109–119, Mar. 2019.
[24] H. Zhang, et al., "Image Prediction for Limited-angle Tomography via Deep Learning with Convolutional Neural Network," *arXiv:1607.08707*, 2016.
[25] D. Wu, K. Kim, G. El Fakhri, and Q. Li, "Iterative low-dose CT reconstruction with priors trained by artificial neural network," *IEEE Trans. Med. Imag.*, vol. 36, no. 12, pp. 2479–2486, Dec. 2017.
[26] Y. Wang, et al., "Iterative quality enhancement via residual-artifact learning networks for low-dose CT," *Phys. Med. Biol.* vol. 63, no. 21, pp. 215004 (17pp), 2018.
[27] K. Gregor and Y. LeCun, "Learning fast approximations of sparse coding," in *Proc. Int. Conf. Mach. Learn. (ICML)*, 2010, pp. 399–406.
[28] S. Roth and M. J. Black, "Fields of experts," *Int. J. Comput. Vis.*, vol. 82, no. 2, pp. 205–229, Apr. 2009.
[29] H. Gupta, K. H. Jin, H. Q. Nguyen, M. T. McCann, and Michael Unser, "CNN-based projected gradient descent for consistent CT image reconstruction," *IEEE Trans. Med. Imag.*, vol. 37, no. 6, pp. 1440–1453, Jun. 2018.
[30] J. Adler and O. Öktem, "Learned primal-dual reconstruction," *IEEE Trans. Med. Imag.*, vol. 37, no. 6, pp. 1322–1332, Jun. 2018.
[31] Y. Yang, J. Sun, H. Li, and Z. Xu, "Deep ADMM-net for compressive sensing MRI," in *Proc. Adv. Neural Inf. Process. Syst. (NIPS)*, 2016, pp. 10–18.
[32] Y. Yang, J. Sun, H. Li, and Z. Xu, "Deep ADMM-net for compressive sensing MRI," *IEEE Trans. Pattern Anal. Mach. Intell.*, vol. 42, no. 3, pp. 521–538, 2020.
[33] J. He et al., "Optimizing a parameterized plug-and-play ADMM for iterative low-dose CT reconstruction," *IEEE Trans. Med. Imag.*, vol. 38, no. 2, pp. 371–382, 2019.
[34] D. Hu, et al., "Hybrid Domain Neural Network Processing for Sparse-view CT Reconstruction," *IEEE Trans. Radiat. Plasma Med. Sci.*, DOI: 10.1109/TRPMS.2020.3011413, 2020.
[35] X. Yin, et al., "Domain Progressive 3D Residual Convolution Network to Improve Low-Dose CT Imaging," *IEEE Trans. Med. Imag.*, vol. 38, no. 12, pp. 2903–2913, 2019.
[36] D. Lee, S, Choi, and H.‐J. Kim, " High quality imaging from sparsely sampled computed tomography data with deep learning and wavelet transform in various domains," *Med. Phys.*, vol. 46, no. 10, pp. 104–115, 2019.
[37] S. Wang et al., "DIMENSION: Dynamic MR imaging with both k-space and spatial prior knowledge obtained via multi-supervised network training," *NMR Biomed.*, DOI: 10.1002/nbm.4131, 2019.
[38] T. Eo, Y. Jun, T. Kim, J. Jang, H. J. Lee, and D. Hwang, "KIKI-net: cross-domain convolutional neural networks for reconstructing undersampled magnetic resonance images," *Magn. Reson. Med.*, vol. 80, no. 5, pp. 2188–2201, 2018.
[39] E. Y. Sidky and X. Pan, "Image reconstruction in circular cone-beam computed tomography by constrained, total-variation minimization," *Phys. Med. Biol.*, vol. 53, no. 17, pp. 4777–4807, 2008.
[40] S. Niu et al., "Sparse-view X-ray CT reconstruction via total generalized variation regularization," *Phys. Med. Biol.*, vol. 59, no. 12, pp. 2997–3017, 2014.
[41] Y. Zhang, W. Zhang, Y. Lei, and J. Zhou, "Few-view image reconstruction with fractional-order total variation," *J. Opt. Soc. Amer. A, Opt. Image Sci.*, vol. 31, no. 5, pp. 981–995, May 2014.
[42] Y. Zhang, Y. Wang, W. Zhang, F. Lin, Y. Pu, and J. Zhou, "Statistical iterative reconstruction using adaptive fractional order regularization," *Biomed. Opt. Exp.*, vol. 7, no. 3, pp. 1015–1029, 2016.
[43] H. Gao, H. Yu, S. Osher, and G. Wang, "Multi-energy CT based on a prior rank, intensity and sparsity model (PRISM)," *Inverse Probl.*, vol. 27, no. 11, 2011, Art. no. 115012.
[44] Y. Lu, J. Zhao, and G. Wang, "Few-view image reconstruction with dual dictionaries," *Phys. Med. Biol.*, vol. 57, no. 1, pp. 173–189, 2012.
[45] Y. Zhang, Y. Xi, Q. Yang, W. Cong, J. Zhou, and G. Wang, "Spectral CT reconstruction with image sparsity and spectral mean," *IEEE Trans. Comput. Imag.*, vol. 2, no. 4, pp. 510–523, Dec. 2016.
[46] J.-F. Cai, X. Jia, H. Gao, S. B. Jiang, Z. Shen, and H. Zhao, "Cine cone beam CT reconstruction using low-rank matrix factorization: algorithm and a proof-of-principle study," *IEEE Trans. Med. Imag.*, vol. 33, no. 8, pp. 1581–1591, Aug. 2014.
[47] W. Xia, et al., "Spectral CT reconstruction – ASSIST: aided by self-similarity in image-spectral tensors," *IEEE Trans. Comput. Imag.*, vol. 5, no. 3, pp. 420–436, 2019.
[48] K. He, X. Zhang, S. Ren, and J. Sun, "Deep residual learning for image recognition," in *Proc. IEEE Conf. Comput. Vis. Pattern Recognit. (CVPR)*, Jun. 2016, pp. 770–778.
[49] H. Shan et al., "3-D convolutional encoder-decoder network for low-dose CT via transfer learning from a 2-D trained network," *IEEE Trans. Med. Imag.*, vol. 37, no. 6, pp. 1522–1534, Jun. 2018.
[50] D. P. Kingma and J. Ba. "Adam: A method for stochastic optimization." arxiv.org/abs/1412.6980, 2014.
[51] R. L. Siddon, "Fast calculation of the exact radiological path for a three-dimensional CT array," *Med. Phys.*, vol. 12, no. 2, pp. 252–255, 1985.
[52] X. Glorot and Y. Bengio, "Understanding the difficulty of training deep feedforward neural networks," in *Proc. Int. Conf. Artif. Intell. Stat. (AISTATS)*, 2010, pp. 249–256.
[53] W. Xia, et al., "CT reconstruction with PDF: parameter-dependent framework for multiple scanning geometries and dose levels," *arXiv:2010.14350*, 2020.